\documentclass[a4paper,11pt]{article}
\usepackage{amssymb}
\usepackage{amsmath}
\usepackage{graphicx}
\usepackage{enumitem}
\usepackage{multirow}
\usepackage[affil-it]{authblk}
\usepackage{caption}
\usepackage{subcaption}
\usepackage{mathtools}
\providecommand{\keywords}[1]
{
	\small    
	\textbf{\textit{Keywords--}} #1
}
\newcommand{\bra}[1]{\ensuremath{\left\langle#1\right|}}
\newcommand{\ket}[1]{\ensuremath{\left|#1\right\rangle}}

\usepackage{tikz}
\usetikzlibrary{matrix,shapes,arrows,positioning,chains, calc}
\newtheorem{example}{\bf Example}

\DeclarePairedDelimiter\floor{\lfloor}{\rfloor}
\begin{document}
	
	\title{Quantum Secure Direct Communication with Mutual Authentication using a Single Basis}
	
	\author{Nayana Das%
		\thanks{Email address: \texttt{dasnayana92@gmail.com} }}
	\affil{Applied Statistics Unit, Indian Statistical Institute, Kolkata, India.}
	
	\author{Goutam Paul %
		\thanks{Email address: \texttt{goutam.paul@isical.ac.in}}}
	\affil{Cryptology and Security Research Unit, R. C. Bose Centre for Cryptology and Security, Indian Statistical Institute, Kolkata, India.}
	
	\author{Ritajit Majumdar %
		\thanks{Email address: \texttt{majumdar.ritajit@gmail.com}}}
	\affil{Advanced Computing \& Microelectronics Unit, Indian Statistical Institute, Kolkata, India.}
	\date{}
	
	\maketitle
	
	\begin{abstract}
		In this paper, we propose a new theoretical scheme for quantum secure direct communication (QSDC) with user authentication. Different from the previous QSDC protocols, the present protocol uses only one orthogonal basis of single-qubit states to encode the secret message. Moreover, this is a one-time and one-way communication protocol, which uses qubits prepared in a randomly chosen arbitrary basis, to transmit the secret message. We discuss the security of the proposed protocol against some common attacks and show that no eavesdropper can get any information from the quantum and classical channels. We have also studied the performance of this protocol under realistic device noise. We have executed the protocol in IBMQ Armonk device and proposed a repetition code based protection scheme that requires minimal overhead.
	\end{abstract}
	
	\keywords{Arbitrary basis; Identity authentication; Quantum cryptography; Secure communication; Without entanglement} 
	
	\section{Introduction}
	\label{intro}
	Nowadays security is one of the basic requirements in our daily life and cryptography is a method of secure communication of our secret information over a public channel. In classical cryptography, there are two types, symmetric or private key cryptography and asymmetric or public key cryptography. Now the security of the asymmetric key cryptosystem is based on some mathematical hardness assumptions, such as integer factorization problem, discrete log problem etc. But due to Shor's algorithm~\cite{shor1994algorithms}, which can factorize an integer in polynomial-time, the quantum computer becomes a threat for asymmetric key cryptography. However, quantum cryptography provides unconditional security based on the fundamental principles of quantum mechanics, such as the Heisenberg uncertainty principle~\cite{heisenberg1985anschaulichen}, quantum no-cloning theory~\cite{wootters1982single}. The concept of quantum cryptography was first introduced by Bennett and Brassard in 1984~\cite{bennett2020quantum} based on the idea of quantum conjugate coding proposed by Wiesner~\cite{wiesner1983conjugate}.
	Since Bennett and Brassard proposed their pioneer work on quantum key distribution (QKD), which is also known as the BB84 protocol~\cite{bennett2020quantum}, a lot of QKD protocols have been presented, such as QKD with entanglement~\cite{ekert1991quantum,long2002theoretically,li2016one}, without entanglement~\cite{bennett1992quantum1,lucamarini2005secure}, experimental QKD~\cite{bennett1992experimental,zhao2006experimental,tang2014experimental,bedington2016nanosatellite,zhong2019proof} and so on.
	
	Different from QKD, QSDC is one of the most important branches of quantum cryptography, which is used to transmit the secret message directly without establishing some prior key for encryption and decryption. Of course, all QSDC protocols can be used for key distribution as it can transmit a predetermined key securely. In the early 2000s, the concept of QSDC was proposed by Long et al.~\cite{long2002theoretically}.  They used Einstein-Podolsky-Rosen (EPR) pairs to transmit the secret message deterministically through the quantum channel. After that, Bostr{\"o}m et al. proposed the famous ping-pong-protocol (PPP) to transfer information in a deterministic secure manner using the EPR pairs~\cite{bostrom2002deterministic}. Later Cai showed that the PPP is insecure against Denial-of-Service (DoS) attack~\cite{cai2004ping}. In 2004 Nguyen improved the PPP and extended it to a bidirectional QSDC protocol, called quantum dialogue, where two legitimate parties can exchange their secret messages simultaneously~\cite{nguyen2004quantum}. Over the past two decades, QSDC has gone through rapid developments~\cite{deng2003two,deng2004secure,wang2005quantum,hu2016experimental,zhang2017quantum,das2020two}. QSDC protocols require higher security than QKD protocols because the secret message is directly transmitted through the quantum channel. Therefore information leakage problem is a serious issue in the direct communication protocols which are briefly discussed in~\cite{zhong2007improvement,gao2008comment,gao2008revisiting,tan2008classical,fei2008teleportation,wang2011information,gao2014information,das2020improving}.

	For secure communication, identity authentication is always important as it prevents an eavesdropper to impersonate a legitimate party. In 1995, Cr{\'e}peau et
	al.~\cite{crepeau1995quantum} proposed the first quantum identification scheme based on quantum oblivious transfer~\cite{bennett1991practical}. QSDC with user authentication was first proposed by Lee et al. in 2006 based on Greenberger-Horne-Zeilinger (GHZ) states~\cite{lee2006quantum}. However, Zhang et al. showed that this protocol is not secure against the intercept-and-resend attack and proposed a revised version of the original protocol~\cite{zhang2007comment}. Later on, a number of new QSDC protocols with authentication are presented~\cite{dan2010new,chang2014controlled,hwang2014quantum,das2020cryptanalysis}.
	
	Almost every quantum cryptographic protocol uses either entangled states or single qubit states randomly prepared in a pair of orthogonal bases, to transmit information securely. In this paper, for the first time, we propose a QSDC protocol, which also provides mutual identity authentication of the participants, by using only one orthogonal basis of single qubit states for encoding the secret message. In the present protocol, the message sender Alice prepares a sequence of single-qubit states corresponding to her message in a randomly chosen arbitrary basis and sends it to the receiver Bob through a quantum channel. Then Alice publicly announces some classical information and they check the security of the channel. If they find any eavesdropper in the channel, then they terminate the protocol. However, in this case the eavesdropper can not get any information about the secret message. After the security check process is passed, then Bob uses the information of Alice to measure the received qubits and to get the secret message. Furthermore, in this protocol, we use only one orthogonal basis to encode all the secret information. But since the basis is chosen arbitrarily, any eavesdropper can not guess the basis of the encoded qubits and therefore the protocol remains secure.
	
	Execution of the protocol in real devices makes them susceptible to the channel noise - in particular decoherence, calibration and readout error. We have executed this protocol in the IBMQ Armonk Device \cite{armonk} to study the behaviour of it in the presence of noise. We show that the effect of noise is equivalent to a bit-flip error in the case of this protocol. We further show from our execution results that the effect of noise does not depend on the choice of basis. In order to account for the non-instantaneous nature of any quantum channel, we model an ideal quantum channel as a series of identity gates without any Eavesdropper. However, in a realistic scenario, these gates are susceptible to noise, and the channel no longer behaves as identity. Our execution results show that a minimal overhead of a 3-qubit repetition code is sufficient to protect this protocol against noise as long as the number of identity gates (i.e. the length of the quantum channel) is below a certain threshold.
	
	The rest of this paper is organized as follows: in Section~\ref{sec2},  we briefly describe our proposed QSDC protocol with an example. In the next section, the security of the protocol is analyzed against all familiar attacks. We next study the effect of noise on this protocol and propose protection schemes against it. Finally Section~\ref{conclusion} concludes our results.
	
	\section{Proposed QSDC protocol with mutual authentication}\label{sec2}
	In this section, we propose the new QSDC protocol with a mutual identity authentication process. We use the basic idea of quantum identity authentication scheme~\cite{ho2017quantum} to verify the identity of the message sender.
	
	Without loss of generality, let Alice be the sender and Bob be the receiver. Also, let Alice and Bob have their previously shared $k$-bit authentication identities (we assume $k$ is even) $Id_A$ and $Id_B$ respectively (using some secured QKD). Alice wants to send a message $M=M_1M_2\ldots M_n$ to Bob. Let $\Theta$ be a predefined set of angles with cardinality $N$. For our protocol, we take $\Theta=\{x^\circ: x $ is an integer and $1 \leq x \leq 360\}$. Thus here, $N=360$. For each $\theta \in \Theta$, the unitary matrix $U_{\theta}$ is defined as
	\begin{equation*}
	U_{\theta}=
	\begin{pmatrix}
	\cos \theta & -\sin \theta\\
	\sin \theta & \cos \theta
	\end{pmatrix}.
	\end{equation*}
	Then $U_{\theta}\ket{0}=\cos \theta \ket{0}+\sin \theta \ket{1}=\ket{x}$ (say), and $U_{\theta}\ket{1}=-\sin \theta \ket{0}+\cos \theta \ket{1}=\ket{y}$ (say).
	The QSDC protocol is as follows:
	
	\begin{enumerate}
		\item Encoding process:
		\begin{enumerate}
			\item Alice puts some random check bits in random positions of her $n$-bit message $M$. Let the new bit string be $M'$, which contains $n'=n+c$ bits, where $c$ is the number of check bits.
			\item She prepares a sequence $Q_A^1$ containing $n'$ number of single qubits in $\{\ket{0}, \ket{1}\}$ basis corresponding to $M'$. She prepares $\ket{0}$ and $\ket{1}$ corresponding to message bit $0$ and $1$ respectively.
			\item Alice randomly chooses an angle $\theta \in \Theta$ and applies the unitary operator $U_{\theta}$ on all the qubits of $Q_A^1$. Thus all the qubits of $Q_A^1$ are now in $\{\ket{x},\ket{y}\}$ basis.
			\item She prepares a sequence of single qubits $I_A$ corresponding to her authentication identity $Id_A$. For $1 \leq i \leq k/2$ (as $k$ is even), she chooses the $i$-th qubit of $I_A$ as $\ket{0},\ket{1},\ket{+}=\frac{1}{\sqrt{2}}(\ket{0}+\ket{1})$ and $\ket{-}=\frac{1}{\sqrt{2}}(\ket{0}-\ket{1})$, according to the values $00,01,10$ and $11$ of the $(2i-1)$-th and the $2i$-th bits of $Id_A$. She randomly inserts the qubits of $I_A$ into $Q_A^1$ and let the new sequence be $Q_A^2$ containing $n'+k/2$ number of qubits.
			\item Alice chooses a $k$-bit random number $r$ and prepares a sequence of single qubits $I_B$ corresponding to the bit strings $Id_B^1=Id_B \oplus r$ and $Id_B$. For $1 \leq i \leq k$, let the $i$-th bit of $Id_B$ ($Id_B^1$) be $Id_{B,i}$ ($Id_{B,i}^1$),
			\begin{enumerate}
				\item if $Id_{B,i}^1=0$ ($1$) and $Id_{B,i} = 0$, then the $i$-th qubit of $I_B$ is $\ket{0}$ ($\ket{1}$),
				\item if $Id_{B,i}^1=0$ ($1$) and $Id_{B,i} = 1$, then the $i$-th qubit of $I_B$ is $\ket{+}$ ($\ket{-}$).
			\end{enumerate}	
			She randomly inserts the qubits of $I_B$ into $Q_A^2$ and let the new sequence be $Q_A^3$ containing $n'+3k/2$ number of qubits.
			\item She also encodes the value of $\theta$ by preparing a sequence of single qubits $Q_\theta$ corresponding to the binary representation of $\theta=\theta_1\theta_2\ldots \theta_{k'}$ containing $k'$ bits. Note that since $\theta$ is an integer, whose value lies between $0$ to $360$, $k' \leq 9$. We assume $k \geq k'$ and then the encoding strategy, for $1 \leq i \leq k'$, is: 
			\begin{enumerate}
				\item if $\theta_i= 0$ ($1$) and $Id_{B,i} = 0$, then prepares $\ket{0}$ ($\ket{1}$),
				\item if $\theta_i= 0$ ($1$) and $Id_{B,i} = 1$, then prepares $\ket{+}$ ($\ket{-}$).
			\end{enumerate}	
			She puts these single qubits in random positions of $Q_A^3$ and let the new sequence be $Q_A^4$ containing $n'+3k/2+k'$ number of qubits.		
			\item Finally she chooses a sequence $D_A$ of $m$ number of decoy photons randomly from $\{\ket{0},\ket{1},$\\$\ket{+},\ket{-}\}$ and inserts them in random positions of $Q_A^4$. Let the new sequence be $Q_A^5$ containing $l=n'+3k/2+k'+m$ single qubits. Alice sends $Q_A^5$ to Bob through a quantum channel.
		\end{enumerate}
		\item Security check: After Bob receives $Q_A^5$, they check if there is any eavesdropper in the channel. Alice announces the positions and bases of the decoy photons. Bob measures the decoy photons and announces the results. By comparing these measurement results and the initial states of the decoy photons, Alice calculates the error in the channel. If the estimated error is greater than some threshold value, then it proves the existence of some eavesdropper in the channel. In that case, they abort the task; otherwise, they continue the protocol.
		\item Authentication procedure: 
		\begin{enumerate}
			\item Alice tells the positions of the single qubits of $I_A$ and Bob measures those qubits in the proper bases corresponding to $Id_A$, i.e., he chooses $\{\ket{0},\ket{1}\} $ basis if the corresponding bits of $Id_A$ are $00$ or $01$; otherwise he chooses  $\{\ket{+},\ket{-}\} $ basis if the corresponding bits of $Id_A$ are $10$ or $11$. Bob compares his measurement results with the bits of $Id_A$ and calculates the error rate. Low error rate implies that there is no eavesdropper impersonating Alice, then he continues the process, otherwise terminates it. 
			
			\item Alice tells the positions of the single qubits of $I_B$ and Bob measures those qubits in the proper bases corresponding to $Id_B$, i.e., he chooses $\{\ket{0},\ket{1}\} $ ($\{\ket{+},\ket{-}\} $) basis if the corresponding bit of $Id_B$ is $0$ ($1$). Then from the measurement results, Bob gets $Id_B^1$ and announces $r=Id_B \oplus Id_B^1$. Alice checks the value of $r$ to confirm Bob's authenticity and decides to continue or abort the communication. 
		\end{enumerate}
		
		\item Decoding process:
		\begin{enumerate}
			\item Alice tells Bob the positions of the qubits of $Q_\theta$ and Bob measures those on proper bases to get the value of $\theta$. Bob discards all the measured qubits and gets back the sequence $Q_A^1$. He applies the unitary operator ${U_\theta}^{-1}$ to all the qubits of $Q_A^1$ and measures these qubits in $\{\ket{0},\ket{1}\}$ basis. If the $i$-th measurement result is $\ket{0}$, then Bob concludes ${M'}_i=0$, else ${M'}_i=1$. 
			\item To check the integrity of the secret message, they publicly compare the random check bits and calculate the error rate. If it is negligible then Bob gets $M$. Otherwise, they abort the protocol.
		\end{enumerate}
	
	\end{enumerate}
	 
	\begin{figure}
		\begin{tikzpicture}
		\matrix (m)[matrix of nodes, column  sep=1cm,row  sep=1mm, nodes={draw=none, text depth=2pt},
		column 1/.style={anchor=base west},
		column 2/.style={anchor=base},
		column 3/.style={anchor=base west} 
		]{
			{\Large Alice (Identity $Id_A$)} & & {\Large Bob (Identity $Id_B$)}\\[2mm]
			\textbf{1. Message Encoding\hfil} & & \\
			Message $M$, chooses $\theta $ and $r$. & & \\
			$\bullet$ Inserts check bits in $M$. & & \\
			$\bullet$ Encodes: $0\rightarrow U_\theta \ket{0},~1\rightarrow U_\theta \ket{1}$. & & \\
			$\bullet$ Prepares sequence $Q_A^1$ & &\\
			$\bullet$ Inserts $I_A,~I_B,~Q_\theta, ~D_A$ in $Q_A^1$. & & \\
			Prepared sequence $Q_A^5$ & Sends  $Q_A^5$ & \\[2mm]
			\textbf{2. Security check} & & \\
			&Position and& Measures qubits of $D_A$.\\
			&bases of $D_A$ &\\[2mm]
			Checks eavesdropping. & Announces & \\
			& states of $D_A$ & \\
			\textbf{3. Authentication process} & & \\
			&Positions of $I_A$ & Measures qubits of $I_A$\\[2mm]
			&  & Checks $Id_A$\\[2mm]
			&Positions of $I_B$ & Calculates $r=Id_B \oplus Id_B^1.$\\[2mm]
			Checks $Id_B$ &Sends $r$ &  \\[2mm]
			\textbf{4. Decoding process} & & \\
			&Positions of $Q_\theta$ & $\bullet$ Measures qubits of $Q_\theta$, gets $\theta$\\
			&& $\bullet$ Discards measured qubits.\\
			&& $\bullet$ Applies ${U_\theta}^{-1}$.\\
			&& $\bullet$ Measures in $\{\ket{0},\ket{1}\}$.\\[2mm]
			&Check bits&  Checks eavesdropping and Get $M$.\\[2mm]
			$\dashrightarrow$ denotes quantum channel&&\\
		    $\longrightarrow$ denotes classical channel&&\\
		    };
		\draw[shorten <=-.1cm,shorten >=-.1cm] (m-1-1.south east)--(m-1-1.south west);
		\draw[shorten <=-.1cm,shorten >=-.1cm] (m-1-3.south east)--(m-1-3.south west);
		\draw[line width = .4mm, dashed,shorten <=-0.4cm,shorten >=-.4cm,-latex] (m-8-2.south west)--(m-8-2.south east);
		\draw[shorten <=-.1cm,shorten >=-.1cm,-latex] (m-10-2.south west)--(m-10-2.south east);
		\draw[shorten <=-.1cm,shorten >=-.1cm,-latex] (m-12-2.south east)--(m-12-2.south west);
		\draw[shorten <=.1cm,shorten >=.1cm,-latex] (m-15-2.south west)--(m-15-2.south east);
		\draw[shorten <=.1cm,shorten >=.1cm,-latex] (m-17-2.south west)--(m-17-2.south east);
		\draw[shorten <=-.4cm,shorten >=-.4cm,-latex] (m-18-2.south east)--(m-18-2.south west);
		\draw[shorten <=.1cm,shorten >=.1cm,-latex] (m-20-2.south west)--(m-20-2.south east);
		\draw[shorten <=-.3cm,shorten >=-.3cm,-latex] (m-24-2.south west)--(m-24-2.south east);
		\draw (current bounding box.north east) -- (current bounding box.north west) -- (current bounding box.south west) -- (current bounding box.south east) -- cycle;
		\end{tikzpicture}
		{\footnotesize \textbf{Notations:} $\theta \in \Theta$, $r \in \{0,1\}^k$,\\
		$I_A:$ qubits corresponding to $Id_A$, $I_B:$ qubits corresponding to $Id_B^1$, $Id_B^1=Id_B \oplus r$,\\
		 $Q_\theta:$ qubits corresponding to $\theta$ and $D_A:$ decoy qubits.}
		\caption{Proposed QSDC protocol with mutual authentication} \label{fig-qsdc}
	\end{figure}

	\begin{example}
		Let us take an example of the above discussed QSDC protocol.\\
		Let $Id_A=1100,~ Id_B=0111$ and the secret message $M=011101$.
		\begin{enumerate}
			\item Encoding process:
			\begin{enumerate}
				\item Alice inserts check bits $1$ and $0$ after the $1$st and $3$rd bits of $M$, i.e., $M'=0\mathbf{1}11\mathbf{0}101.$ (Bold numbers are check bits.)
				\item $Q^1_A=\ket{0}\ket{1}\ket{1}\ket{1}\ket{0}\ket{1}\ket{0}\ket{1}$.
				\item Alice chooses $\theta=7^\circ$ and applies $U_\theta$ on the qubits of $Q^1_A$. Then $Q^1_A=\ket{x}\ket{y}\ket{y}\ket{y}\ket{x}\ket{y}\ket{x}\ket{y}$, where $\ket{x}=U_\theta\ket{0}$, $\ket{y}=U_\theta\ket{1}$.
				\item $I_A=\ket{-}\ket{0}$ and $Q^2_A=\ket{x}\ket{y}$
				\begin{tikzpicture}[baseline=(char.base)]
				\node(char)[draw,fill=white,
				shape=rectangle,
				minimum width=.1cm]
				{$\mathbf{\ket{-}}$};
				\end{tikzpicture}
				$\ket{y}$
				\begin{tikzpicture}[baseline=(char.base)]
				\node(char)[draw,fill=white,
				shape=rectangle,
				minimum width=.1cm]
				{$\ket{0}$};
				\end{tikzpicture}
				$\ket{y}\ket{x}\ket{y}\ket{x}\ket{y}$, where the boxed qubits are randomly added from $I_A$. 
				
				\item Alice chooses $r=1001$, then $Id_B^1=Id_B \oplus r=0111 \oplus 1001= 1110,~I_B=\ket{1}\ket{-}\ket{-}\ket{+}$ and $Q^3_A=\ket{x}$
				\begin{tikzpicture}[baseline=(char.base)]
				\node(char)[draw,fill=white,
				shape=rectangle,
				minimum width=.1cm]
				{$\ket{1}$};
				\end{tikzpicture}
				$\ket{y}\ket{-}$
				\begin{tikzpicture}[baseline=(char.base)]
				\node(char)[draw,fill=white,
				shape=rectangle,
				minimum width=.1cm]
				{$\ket{-}$};
				\end{tikzpicture}
				$\ket{y}\ket{0}\ket{y}$
				\begin{tikzpicture}[baseline=(char.base)]
				\node(char)[draw,fill=white,
				shape=rectangle,
				minimum width=.1cm]
				{$\mathbf{\ket{-}}$};
				\end{tikzpicture}
				$\ket{x}\ket{y}\ket{x}$
				\begin{tikzpicture}[baseline=(char.base)]
				\node(char)[draw,fill=white,
				shape=rectangle,
				minimum width=.1cm]
				{$\mathbf{\ket{+}}$};
				\end{tikzpicture}
				$\ket{y}$, where the boxed qubits are randomly added from $I_B$.
				\item $Q_\theta=\ket{1}\ket{-}\ket{-}$  and $Q^4_A=\ket{x}\ket{1}\ket{y}\ket{-}\ket{-}\ket{y}$
				\begin{tikzpicture}[baseline=(char.base)]
				\node(char)[draw,fill=white,
				shape=rectangle,
				minimum width=.1cm]
				{$\ket{1}$};
				\end{tikzpicture}
				$\ket{0}$
				\begin{tikzpicture}[baseline=(char.base)]
				\node(char)[draw,fill=white,
				shape=rectangle,
				minimum width=.1cm]
				{$\ket{-}$};
				\end{tikzpicture}
				$\ket{y}\ket{-}\ket{x}\ket{y}$
				\begin{tikzpicture}[baseline=(char.base)]
				\node(char)[draw,fill=white,
				shape=rectangle,
				minimum width=.1cm]
				{$\ket{-}$};
				\end{tikzpicture}
				$\ket{x}\ket{+}\ket{y}$, where the boxed qubits are randomly added from $Q_\theta$.
				\item Decoy photons $D_A=\ket{0}\ket{1}\ket{+}\ket{0}$ and $Q^5_A=\ket{x}$
				\begin{tikzpicture}[baseline=(char.base)]
				\node(char)[draw,fill=white,
				shape=rectangle,
				minimum width=.1cm]
				{$\ket{0}$};
				\end{tikzpicture}
				$\ket{1}$
				\begin{tikzpicture}[baseline=(char.base)]
				\node(char)[draw,fill=white,
				shape=rectangle,
				minimum width=.1cm]
				{$\ket{1}$};
				\end{tikzpicture}
				$\ket{y}\ket{-}\ket{-}\ket{y}\ket{1}\ket{0}\ket{-}\ket{y}\ket{-}\ket{x}$
				\begin{tikzpicture}[baseline=(char.base)]
				\node(char)[draw,fill=white,
				shape=rectangle,
				minimum width=.1cm]
				{$\ket{+}$};
				\end{tikzpicture}
				$\ket{y}\ket{-}\ket{x}$
				\begin{tikzpicture}[baseline=(char.base)]
				\node(char)[draw,fill=white,
				shape=rectangle,
				minimum width=.1cm]
				{$\ket{0}$};
				\end{tikzpicture}
				$\ket{+}\ket{y}$, where the boxed qubits are randomly added from $D_A$. 
				\item Alice sends $Q_A^5=\ket{x}\ket{0}\ket{1}\ket{1}\ket{y}\ket{-}\ket{-}\ket{y}\ket{1}\ket{0}\ket{-}\ket{y}\ket{-}\ket{x}\ket{+}\ket{y}\ket{-}\ket{x}\ket{0}\ket{+}\ket{y}$ to Bob.	
			\end{enumerate}
			
			\item Security check: After Bob receives $Q_A^5$, Alice announces the positions $(2$nd, $4$th, $15$th and $19$th$)$ and bases $(\{\ket{0},\ket{1}\}$, $\{\ket{0},\ket{1}\}$, $\{\ket{+},\ket{-}\}$, $\{\ket{0},\ket{1}\})$ of the decoy photons. Bob measures the decoy photons and announces the results $(\ket{0},\ket{1},\ket{+},\ket{0})$. Alice calculates the error in the channel. Here, we assume a noiseless channel. Hence, Bob discards all the measured qubits and gets back the sequence $Q_A^4$.
			
			\item Authentication procedure: 
			\begin{enumerate}
				\item Alice announces the positions $(4$th and $8$th $)$ of the qubits of $I_A$ and Bob chooses the bases $(\{\ket{+},\ket{-}\}$, $\{\ket{0},\ket{1}\}$ to measure those qubits and gets $\ket{-}\ket{0}$, which is equivalent to $Id_A$. 
								
				\item Alice tells the positions $(2$nd, $5$th, $11$th and $16$th$)$ of the single qubits of $I_B$ and Bob chooses the bases $(\{\ket{0},\ket{1}\}$, $\{\ket{+},\ket{-}\}$,  $\{\ket{+},\ket{-}\})$ and $\{\ket{+},\ket{-}\})$ to measure those qubits and gets $\ket{1}\ket{-}\ket{-}\ket{+}$.
				He gets $Id_B^1=1110$ announces $r=1110 \oplus 0111=1001$. Alice confirms Bob's identity.
			\end{enumerate}
			
			\item Decoding process:
			\begin{enumerate}
				\item Alice tells Bob the positions $(7$th, $9$th and $14$th$)$ of the qubits of $Q_\theta$ and Bob chooses the bases $(\{\ket{0},\ket{1}\}$, $\{\ket{+},\ket{-}\}$,  $\{\ket{+},\ket{-}\})$ to measure those qubits and obtains $\theta$.
				\item He discards all the measured qubits to get $Q_A^1$ and applies  ${U_\theta}^{-1}$ to all the qubits of $Q_A^1$. Bob measures these qubits in $\{\ket{0},\ket{1}\}$ basis and gets $M'=01110101.$ 
				\item They publicly compare the random check bits $(2$nd and $5$th bit of $M')$ and Bob discards those bits to obtain $ M=011101$. 
			\end{enumerate}
		\end{enumerate}
		This completes the QSDC protocol.		
		\end{example}
		
	\section{Security analysis}
	We now discuss the security of the proposed protocol against some familiar attack strategies such as the impersonation attack, intercept-and-resend attack, entangle-and-measure attack, denial-of-Service (DoS) attack, man-in-the-middle attack, information leakage attack, and Trojan horse attack.   
	\begin{enumerate}
		\item \textbf{Impersonation attack:} Let us first discuss this attack model, where an eavesdropper (Eve) is impersonating a legitimate party. 		
		First, we assume Eve impersonates Alice to send a wrong message to Bob. Since Eve has no knowledge about $Id_A$, she prepares the qubits of $I_{A}'$ randomly from $\{\ket{0},\ket{1},\ket{+},\ket{-}\}$. As Bob knows $Id_A$, he chooses the corresponding bases to measure the qubits of $I_{A}'$. According to the value of the bits $Id_{A,(2i-1)}Id_{A,2i}$, let the $i$-th qubit of $I_{A}$ be $I_{A,i}$ prepared in basis $\mathcal{B}$, where $\mathcal{B}=\{\ket{0},\ket{1}\}$ or $\{\ket{+},\ket{-}\}$. Also let Eve prepare the $i$-th qubit $I_{A,i}'$ in $\mathcal{B}'$ basis. Since Bob knows the exact state of $I_{A,i}$, he measures $I_{A,i}'$ in $\mathcal{B}$ basis and let the measurement result be $I_{A,i}''$. Now the probability that Bob can not find this eavesdropping is $\Pr(I_{A,i}''= I_{A,i})$. Now,
		\begin{itemize}
			\item If $\mathcal{B} = \mathcal{B}'$ and $I_{A,i}=I_{A,i}'$, then $I_{A,i}''=I_{A,i}$ with probability $1$.
			\item If $\mathcal{B} = \mathcal{B}'$ and $I_{A,i} \neq I_{A,i}'$, then $I_{A,i}''=I_{A,i}$ with probability $0$.
			\item If $\mathcal{B} \neq \mathcal{B}'$, then $I_{A,i}''= I_{A,i}$ with probability $1/2$.
		\end{itemize}
		Thus for each qubit of $I_{A}'$ the winning probability of Eve is
		\begin{equation*} \label{eq-pr}
		\begin{split}
		 &\Pr(I_{A,i}''=I_{A,i}) \\
		& =  \Pr(I_{A,i}''=I_{A,i}|~\mathcal{B} = \mathcal{B}')\Pr(\mathcal{B} = \mathcal{B}') + \Pr(I_{A,i}''=I_{A,i}|~\mathcal{B} \neq \mathcal{B}')\Pr(\mathcal{B} \neq \mathcal{B}') \\
		&= \frac{1}{2}[\Pr(I_{A,i}''=I_{A,i}|~\mathcal{B} = \mathcal{B}') + \Pr(I_{A,i}''=I_{A,i}|~\mathcal{B} \neq \mathcal{B}')] \\
		&=  \frac{1}{2}[\Pr(I_{A,i}''=I_{A,i}|~\mathcal{B} = \mathcal{B}',~I_{A,i}=I_{A,i}') \Pr(I_{A,i}=I_{A,i}') + \\
		&~~~~~~~~~~~~~~ \Pr(I_{A,i}''=I_{A,i}|~\mathcal{B} = \mathcal{B}',~I_{A,i} \neq I_{A,i}') \Pr(I_{A,i} \neq I_{A,i}') +1/2]\\
		& = \frac{1}{2}\left[1 \times \frac{1}{2} + 0 \times \frac{1}{2} + \frac{1}{2}\right]=\frac{1}{2}.
		\end{split}
		\end{equation*}
		Hence in the authentication process, Bob can detect Eve with probability $1-(1/2)^{k/2}$. 
		
		On the other hand, now let Eve impersonate Bob to get the secret message from Alice. Then Eve has no idea about the preparation bases of the qubits of $I_B$ and thus she randomly chooses basis $\{\ket{0},\ket{1}\}$ or $\{\ket{+},\ket{-}\}$ to measure those qubits. From the measurement results, she correctly guesses the value of $Id_B^1$ with probability $(3/4)^k$. Since $Id_B^1=Id_B \oplus r$ and $Id_B$ is unknown to Eve, from the security notion of ``One-Time-Pad'', $r$ is completely random to her and she correctly guesses $r$ with probability $(1/2)^k$. 
		Therefore, when Eve announces the random number $r$, Alice detects her with probability $1-(1/2)^k$.
		
		So for both cases, the legitimate party can detect the eavesdropping with a high probability.	
		
		\item \textbf{Intercept-and-resend attack:} 
		In this attack model, Eve intercepts the qubits from the quantum channel from Alice to Bob, then she measures those qubits and resends to Bob. In our proposed protocol, let Eve intercept the sequence $Q_A^5$ from the quantum channel. Note that the qubits corresponding to $M'$ are encoded in an arbitrary basis $\{\ket{x},\ket{y}\}$ and those are in random positions of $Q_A^5$. Let Eve choose a random $\theta_0 \in \Theta$ and measure all the qubits in $\{\ket{x_0},\ket{y_0}\}$ basis, where, 
		
		\begin{equation} \label{eq1}
		\begin{split}
		\ket{x_0}=U_{\theta_0}\ket{0} & = \cos \theta_0 \ket{0}+\sin \theta_0 \ket{1} \\
		& = \frac{1}{\sqrt{2}}[(\cos \theta_0+\sin \theta_0)\ket{+}+(\cos \theta_0 - \sin \theta_0)\ket{-}]
		\end{split}
		\end{equation}
		and 
		\begin{equation} \label{eq2}
		\begin{split}
		\ket{y_0}=U_{\theta_0}\ket{1} & = -\sin \theta_0 \ket{0}+\cos \theta_0 \ket{1} \\
		& = \frac{1}{\sqrt{2}}[(\cos \theta_0 - \sin \theta_0)\ket{+} - (\cos \theta_0 + \sin \theta_0)\ket{-}].
		\end{split}
		\end{equation}
		Then, 
		\begin{equation} \label{eq3}
		\begin{split}
		\ket{0} & = \cos \theta_0 \ket{x} -\sin \theta_0 \ket{y}, \\
		\ket{1} & = \sin \theta_0 \ket{x} +\cos \theta_0 \ket{y}
		\end{split}
		\end{equation}
		and 
		\begin{equation} \label{eq4}
		\begin{split}
		\ket{+}& = \frac{1}{\sqrt{2}}[(\cos \theta_0 + \sin \theta_0)\ket{x} + (\cos \theta_0 - \sin \theta_0)\ket{y}], \\
		\ket{-} & = \frac{1}{\sqrt{2}}[(\cos \theta_0 - \sin \theta_0)\ket{x} - (\cos \theta_0 + \sin \theta_0)\ket{y}].
		\end{split}
		\end{equation}

		\begin{table}[h]
		\centering
			\caption{Effects of Eve's measurement on decoy photons}
			\begin{tabular}{|c|c|c|c|c|}
				\hline
				& \multicolumn{2}{c|}{\begin{tabular}[c]{@{}c@{}}\textbf{After Eve's}\\ \textbf{measurement: $D_{A,i}'$}\end{tabular}}                    & \multicolumn{2}{c|}{\begin{tabular}[c]{@{}c@{}}\textbf{After Bob's}\\ \textbf{measurement: $D_{A,i}''$}\end{tabular}}            \\ \cline{2-5} 
				\multirow{-2}{*}{\begin{tabular}[c]{@{}c@{}}\textbf{Original}\\ \textbf{state} $D_{A,i}$\end{tabular}} & \textbf{State}                      & \textbf{Probability}                           & \textbf{State}              & \textbf{Probability}                           \\ \hline
				& $\ket{x_0}$                         & ${\cos^2 \theta_0}$                            &                             & ${\cos^2 \theta_0}$                            \\ \cline{2-3} \cline{5-5} 
				\multirow{-2}{*}{$\ket{0}$}                         & $\ket{y_0}$                         & ${\sin^2 \theta_0}$                            & \multirow{-2}{*}{$\ket{0}$} & ${\sin^2 \theta_0}$                            \\ \hline
				& $\ket{x_0}$                         & ${\sin^2 \theta_0}$                            &                             & ${\sin^2 \theta_0}$                            \\ \cline{2-3} \cline{5-5} 
				\multirow{-2}{*}{$\ket{1}$}                         & $\ket{y_0}$                         & ${\cos^2 \theta_0}$                            & \multirow{-2}{*}{$\ket{1}$} & ${\cos^2 \theta_0}$                            \\ \hline
				& $\ket{x_0}$                         & $\frac{1}{2}(\cos \theta_0 + \sin \theta_0)^2$ &                             & $\frac{1}{2}(\cos \theta_0 + \sin \theta_0)^2$ \\ \cline{2-3} \cline{5-5} 
				\multirow{-2}{*}{$\ket{+}$}                         & $\ket{y_0}$                         & $\frac{1}{2}(\cos \theta_0 - \sin \theta_0)^2$ & \multirow{-2}{*}{$\ket{+}$} & $\frac{1}{2}(\cos \theta_0 - \sin \theta_0)^2$ \\ \hline
				& $\ket{x_0}$ & $\frac{1}{2}(\cos \theta_0 - \sin \theta_0)^2$ &                             & $\frac{1}{2}(\cos \theta_0 - \sin \theta_0)^2$ \\ \cline{2-3} \cline{5-5} 
				\multirow{-2}{*}{$\ket{-}$}                         & $\ket{y_0}$                         & $\frac{1}{2}(\cos \theta_0 + \sin \theta_0)^2$ & \multirow{-2}{*}{$\ket{-}$} & $\frac{1}{2}(\cos \theta_0 +\sin \theta_0)^2$  \\ \hline
			\end{tabular}
			\label{table}
		\end{table}
		
		Eve's measurement affects the decoy photons as well. Let the $i$-th decoy photon be $D_{A,i}$ prepared in basis $\mathcal{B}$, where $\mathcal{B}=\{\ket{0},\ket{1}\}$ or $\{\ket{+},\ket{-}\}$, and after Eve measures in $\{\ket{x_0},\ket{y_0}\}$ basis the state becomes $D_{A,i}'$. When Alice announces the preparation basis of $D_{A,i}$, then Bob measures $D_{A,i}'$ in basis $\mathcal{B}$ and gets $D_{A,i}''$. We now calculate the probability that $D_{A,i}=D_{A,i}''$. From Table~\ref{table} we get,
		\begin{equation*} \label{eq-pr1}
		\begin{split}
		&\Pr(D_{A,i}''=D_{A,i}) \\
		& =  \sum_{\ket{b} \in \{\ket{0},\ket{1}\}} \Pr(D_{A,i}''=\ket{b},D_{A,i}=\ket{b})+ \sum_{\ket{b} \in \{\ket{+},\ket{-}\}} \Pr(D_{A,i}''=\ket{b},D_{A,i}=\ket{b}) \\
		&= \sum_{\ket{b} \in \{\ket{0},\ket{1}\}} \Pr(D_{A,i}''=\ket{b}|~D_{A,i}=\ket{b})\Pr(D_{A,i}=\ket{b})+ \\
		&~~~~~~~~~\sum_{\ket{b} \in \{\ket{+},\ket{-}\}} \Pr(D_{A,i}=\ket{b}|~D_{A,i}''=\ket{b})\Pr(D_{A,i}=\ket{b}) \\
		& = \frac{1}{4}\left[ \sum_{\ket{b} \in \{\ket{0},\ket{1}\}} \Pr(D_{A,i}''=\ket{b}|~D_{A,i}=\ket{b})+ \sum_{\ket{b} \in \{\ket{+},\ket{-}\}} \Pr(D_{A,i}''=\ket{b}|~D_{A,i}=\ket{b})\right] \\
		& = \frac{1}{4}\left[2\left(\cos^4 \theta_0+\sin^4 \theta_0\right) + 2\left\lbrace  \frac{1}{4}\left(\cos \theta_0+\sin \theta_0\right)^4+ \frac{1}{4}\left(\cos \theta_0 - \sin \theta_0\right)^4 \right\rbrace   \right]\\
		& = \frac{1}{2}\left[\left(\cos^4 \theta_0+\sin^4 \theta_0\right) +  \frac{1}{2}\left(1+sin ^2 2\theta_0  \right)   \right]\\
		& = \frac{1}{2}\left(sin^2 \theta_0 +cos^2 \theta_0 \right)^2 + \frac{1}{4}=  \frac{3}{4}.
		\end{split}
		\end{equation*}
		Thus the probability that Alice and Bob can realize the existence of Eve is 
		$1-\left( \frac{3}{4}\right) ^m$, where $m$ is the number of decoy photons. However, in this case the legitimate parties detect her and terminates the protocol.   
		
		Now, let us calculate the probability $p_{corr}$, that Eve guesses the original $n$-bit message $M$ of Alice correctly. If Eve chooses $\theta_0=\theta$ and measures the qubits of the sequence $Q_A^5$ in $\{\ket{x},\ket{y}\}$ basis, then she have to choose the correct $n$ positions corresponding to the message bits among $l=n'+3k/2+k'+m$ positions. Thus the winning probability of Eve is:
		$$p_{corr}=\frac{1}{N \times {l \choose n}}.$$
		For positive integers $n$ and $l$ with $1 \leq n \leq l$, we know that, ${\left(\frac{l}{n}\right)}^n \leq {l \choose n}$, which implies $$p_{corr} \leq \frac{1}{N}{\left(\frac{n}{l}\right)}^n \leq \left(\frac{1}{2}\right)^{\floor*{log_2 N}} \times {\left(\frac{n}{l}\right )}^n \leq \left(\frac{1}{2}\right)^n, \text{ if } l \geq 2n\left(\frac{1}{2}\right)^{{\floor*{log_2 N}}/n},$$
		where ${\floor*{log_2 N}}$ denotes the greatest integer less than or equal to $log_2 N$. So for our case $p_{corr} \leq \left(\frac{1}{2}\right)^n$, if $l \geq 2n\left(\frac{1}{2}\right)^{8/n}$.
		Since $p_{corr}$ is negligible, our protocol is secure against this attack strategy.

		\item \textbf{Entangle-and-measure attack:}
		In addition to the above discussed attacks, there is a different kind of attack, called entangle-and-measure attack, which Eve can apply to get a partial information about $M$. For this purpose, Eve prepares a set of ancilla qubits whose initial states are $\ket{\chi}_e$. When Alice sends $Q_A^5$ to Bob, Eve performs a unitary operation $\mathcal{U}_e$ on the qubits of $Q_A^5$ and $\ket{\chi}_e$ to make them entangled, where $\mathcal{U}_e$ is defined as~\cite{gisin2002quantum}:
		\begin{equation}\label{ent}
		\begin{aligned}
		\mathcal{U}_e\ket{0}\ket{\chi}_e=\alpha_0 \ket{0}\ket{\chi_{00}}_e+ \beta_0 \ket{1}\ket{\chi_{01}}_e,\\
		\mathcal{U}_e\ket{1}\ket{\chi}_e=\alpha_1 \ket{0}\ket{\chi_{10}}_e+ \beta_1 \ket{1}\ket{\chi_{11}}_e,
		\end{aligned}
		\end{equation}
		where the four pure states $\ket{\chi_{00}}_e,~\ket{\chi_{01}}_e,~\ket{\chi_{10}}_e$ and $\ket{\chi_{11}}_e$ are orthonormal and they belong to Eve's Hilbert space. They are uniquely determined by the unitary operation $\mathcal{U}_e$ and the following conditions hold,
		\begin{equation}
		\begin{aligned}
		|\alpha_0|^2+|\beta_0|^2=1,~|\alpha_1|^2+|\beta_1|^2=1,\\
		|\alpha_0|^2=|\beta_1|^2=\mathcal{F},~|\alpha_1|^2=|\beta_0|^2=\mathcal{D}. \\
		\end{aligned}
		\end{equation}
		If Alice sends $\ket{b}$, $b\in \{0,1\}$, then after measurement Bob gets the correct result with probability $\mathcal{F}$. Here $\mathcal{F}$ is the fidelity and $\mathcal{D}$ is the quantum bit error rate (QBER). 
		
		Further, we get
		\begin{equation} \label{ent-atck}
		\begin{split}
		\mathcal{U}_e\ket{+}\ket{\chi}_e& = \frac{1}{\sqrt{2}}\left(\mathcal{U}_e\ket{0}\ket{\chi}_e +  \mathcal{U}_e\ket{1}\ket{\chi}_e\right) \\
		&= \frac{1}{\sqrt{2}}\left[ \alpha_0 \ket{0}\ket{\chi_{00}}_e+ \beta_0 \ket{1}\ket{\chi_{01}}_e + \alpha_1 \ket{0}\ket{\chi_{10}}_e+ \beta_1 \ket{1}\ket{\chi_{11}}_e\right] \\
		& = \frac{1}{\sqrt{2}}[ ~\ket{+}( \alpha_0 \ket{\chi_{00}}_e + \beta_0 \ket{\chi_{01}}_e + \alpha_1 \ket{\chi_{10}}_e + \beta_1 \ket{\chi_{11}}_e)/\sqrt{2}~ +\\
		&~~~~~~ \ket{-}( \alpha_0 \ket{\chi_{00}}_e - \beta_0 \ket{\chi_{01}}_e + \alpha_1 \ket{\chi_{10}}_e - \beta_1 \ket{\chi_{11}}_e )/\sqrt{2}~] \\
		& = \frac{1}{\sqrt{2}}(\ket{+}\ket{\chi_{++}}_e+\ket{-}\ket{\chi_{+-}}_e)
		\end{split}
		\end{equation}
		and
		\begin{equation} \label{ent-atck1}
		\begin{split}
		\mathcal{U}_e\ket{-}\ket{\chi}_e& = \frac{1}{\sqrt{2}}\left(\mathcal{U}_e\ket{0}\ket{\chi}_e -  \mathcal{U}_e\ket{1}\ket{\chi}_e\right) \\
		&= \frac{1}{\sqrt{2}}\left[ \alpha_0 \ket{0}\ket{\chi_{00}}_e+ \beta_0 \ket{1}\ket{\chi_{01}}_e - \alpha_1 \ket{0}\ket{\chi_{10}}_e - \beta_1 \ket{1}\ket{\chi_{11}}_e\right] \\
		& = \frac{1}{\sqrt{2}}[ ~\ket{+}( \alpha_0 \ket{\chi_{00}}_e + \beta_0 \ket{\chi_{01}}_e - \alpha_1 \ket{\chi_{10}}_e - \beta_1 \ket{\chi_{11}}_e)/\sqrt{2}~ +\\
		&~~~~~~ \ket{-}( \alpha_0 \ket{\chi_{00}}_e - \beta_0 \ket{\chi_{01}}_e - \alpha_1 \ket{\chi_{10}}_e + \beta_1 \ket{\chi_{11}}_e )/\sqrt{2}~] \\
		& = \frac{1}{\sqrt{2}}(\ket{+}\ket{\chi_{-+}}_e+\ket{-}\ket{\chi_{--}}_e).
		\end{split}
		\end{equation}
		If Alice sends $\ket{b}$, $b\in \{+,-\}$, then after measurement Bob gets the correct result with probability $1/2$.
		
		Now in the present protocol Alice prepares decoy states randomly from $\{\ket{0},\ket{1},\ket{+},\ket{-}\}$. So for a particular decoy state $\ket{b}$, Bob gets the correct state with probability $p=\frac{1}{2}(\mathcal{F}+1/2)$, where $\mathcal{F}$ is the fidelity when the decoy state is in $\{\ket{0},\ket{1}\}$ and $1/2$ is the fidelity when the decoy state is in $\{\ket{+},\ket{-}\}$. Moreover, both of these cases occur with probability $1/2$. Hence in security check Alice and Bob can detect Eve with probability $1-{p}^m$, where $m$ is the number of decoy states.
		
		However we now show that, by applying this attack strategy, Eve gets no information about the secret message. From Equation~\eqref{ent} we have, 
		\begin{equation} \label{ent-atck2}
		\begin{split}
		\mathcal{U}_e\ket{x}\ket{\chi}_e& = \mathcal{U}_e(cos \theta \ket{0} + \sin\theta \ket{1})\ket{\chi}_e \\
		&= \ket{0}(\alpha_0\cos \theta \ket{\chi_{00}}_e + \alpha_1\sin\theta \ket{\chi_{10}}_e) + \ket{1}( \beta_0\cos \theta \ket{\chi_{01}}_e + \beta_1\sin\theta \ket{\chi_{11}}_e) \\
		& =(\cos \theta \ket{x} -\sin \theta \ket{y})(\alpha_0\cos \theta \ket{\chi_{00}}_e + \alpha_1\sin\theta \ket{\chi_{10}}_e)+\\
		&~~~~~~ (\sin \theta \ket{x} +\cos \theta \ket{y})( \beta_0\cos \theta \ket{\chi_{01}}_e + \beta_1\sin\theta \ket{\chi_{11}}_e)
		\end{split}
		\end{equation}
		and 
		\begin{equation} \label{ent-atck3}
		\begin{split}
		\mathcal{U}_e\ket{y}\ket{\chi}_e& = \mathcal{U}_e(- \sin \theta \ket{0} + \cos \theta \ket{1})\ket{\chi}_e \\
		&= \ket{0}(-\alpha_0\sin\theta \ket{\chi_{00}}_e + \alpha_1 \cos \theta \ket{\chi_{10}}_e) + \ket{1}( -\beta_0 \sin \theta \ket{\chi_{01}}_e + \beta_1 \cos \theta \ket{\chi_{11}}_e) \\
		& =(\cos \theta \ket{x} -\sin \theta \ket{y})(-\alpha_0\sin\theta \ket{\chi_{00}}_e + \alpha_1 \cos \theta \ket{\chi_{10}}_e)+\\
		&~~~~~~ (\sin \theta \ket{x} +\cos \theta \ket{y})( -\beta_0 \sin \theta \ket{\chi_{01}}_e + \beta_1 \cos \theta \ket{\chi_{11}}_e).
		\end{split}
		\end{equation}
		From Equation \eqref{ent-atck2} and \eqref{ent-atck3} it follows that, Eve gains no useful information by measuring the ancilla qubit $\ket{\chi}_e$ entangled with the qubits corresponding to the secret message. 
		
		\item \textbf{DoS attack:}
		In this attack model, Eve's aim is not to get secret information but to tamper with the original message~\cite{cai2004ping}. To execute this attack strategy, Eve intercepts the qubits from the quantum channel and randomly applies $I$ and $U$ with probability $1/2$, where $U$ is a random unitary operator. Since Eve does not know the positions of the decoy state, the unitary operation also affects those qubits. 
		
		As the Pauli matrices~\cite{nielsen2002quantum} $I$, $\sigma_x$, $i\sigma_y$ and $\sigma_{z}$ form a basis for the space of all $2 \times 2$ Hermitian matrices, thus the 
		unitary matrix $U$ can be represented as a linear combination of the Pauli matrices. Let 
		$$U=w_1 I + w_2 \sigma_x +i w_3 \sigma_y+ w_4 \sigma_{z},$$
		since $U$ is unitary, we must have $\sum _{i=1}^4 w_i^2=1$, we consider only real coefficients. To calculate the winning probability of Eve, let us first discuss the effects of the Pauli operators on the decoy qubits.
		
		$I$ is the identity operator, so it does not change the state of any qubit. Hence if Eve applies $I$ on a decoy state, then after measurement Bob gets the correct result with probability $p_1=1$.		
		\begin{equation}
		\begin{split}
		\sigma_x \ket{0}=\ket{1},~\sigma_x \ket{1}=\ket{0},~
		\sigma_x \ket{+}=\ket{+},~\sigma_x \ket{-}=-\ket{-},
		\end{split}
		\end{equation}
		i.e., if Eve applies $\sigma_x$ on a decoy state, then after measurement Bob gets the correct result with probability $p_2=1/2$, as $\sigma_x$ changes the state of a decoy qubit $\ket{d}$ only if $\ket{d} \in \{\ket{0},\ket{1}\}$.
		
		Similarly,
		\begin{equation}
		\begin{split}
		i\sigma_y \ket{0}=-\ket{1},~i\sigma_y \ket{1}=\ket{0},~
		i\sigma_y \ket{+}=\ket{-},~i\sigma_y \ket{-}=-\ket{+},
		\end{split}
		\end{equation}
		and 
		\begin{equation}
		\begin{split}
		\sigma_z \ket{0}=\ket{0},~\sigma_z \ket{1}=-\ket{1},~
		\sigma_z \ket{+}=\ket{-},~\sigma_z \ket{-}=\ket{+},
		\end{split}
		\end{equation}
		i.e., if Eve applies $i\sigma_y$ (or $\sigma_z$) on a decoy state, then after measurement Bob gets the correct result with probability $p_3=0$ (or $p_4=1/2$). Thus when Eve applies $U$ on the decoy qubits, then the winning probability of Eve is $$p'=\sum _{i=1}^4 p_i w_i^2 < 1 \text{ as } U\neq I.$$		
		
		Now Eve chooses $I$ and $U$ with probability $1/2$ and thus the probability that Bob gets the correct result is $p''=(1+p')/2$. Hence in the security check process Alice and Bob find this eavesdropping with probability $1-{p''}^{m}>0$, where $m$ is the number of decoy states. Moreover, this attack can also be found when they publicly compare the random check bits to check the integrity of the message.
		
		\item \textbf{Man-in-the-middle attack:} 
		When Eve follows this attack strategy, she intercepts the sequence $Q_A^5$ from the quantum channel and keeps this. She prepares another set $Q_E$ of single qubit states and sends $Q_E$ to Bob instead of $Q_A^5$. Since Eve does not know the position and exact states of the decoy qubits, she prepares all the single qubits in $\{\ket{0},\ket{1}\}$ and $\{\ket{+},\ket{-}\}$ bases to reduce the detection probability in the security check process. Let the $i$-th decoy photon be $D_{A,i}$, which is the $j$-th qubit of the sequence $Q_A^5$, prepared in basis $\mathcal{B}$. Also let the $j$-th qubit of $Q_E$ be $D_{A,i}'$ prepared in basis $\mathcal{B}'$, where $\mathcal{B} \text{ and } \mathcal{B}' \text{ are } \{\ket{0},\ket{1}\}$ or $\{\ket{+},\ket{-}\}$. In the security check process when Alice announces the preparation basis of $D_{A,i}$, then Bob measures $D_{A,i}'$ in basis $\mathcal{B}$ and gets $D_{A,i}''$. We now calculate the probability that $D_{A,i}''=D_{A,i}$. 
		\begin{itemize}
			\item If $\mathcal{B} = \mathcal{B}'$ and $D_{A,i}=D_{A,i}'$, then $D_{A,i}''=D_{A,i}$ with probability $1$.
			\item If $\mathcal{B} = \mathcal{B}'$ and $D_{A,i} \neq D_{A,i}'$, then $D_{A,i}''=D_{A,i}$ with probability $0$.
			\item If $\mathcal{B} \neq \mathcal{B}'$, then $D_{A,i}''=D_{A,i}$ with probability $1/2$.
		\end{itemize}
		Thus for each decoy qubit, the winning probability of Eve is
		\begin{equation*} \label{eq-pr}
		\begin{split}
		&\Pr(D_{A,i}''=D_{A,i}) \\
		& =  \Pr(D_{A,i}''=D_{A,i}|~\mathcal{B} = \mathcal{B}')\Pr(\mathcal{B} = \mathcal{B}') + \Pr(D_{A,i}''=D_{A,i}|~\mathcal{B} \neq \mathcal{B}')\Pr(\mathcal{B} \neq \mathcal{B}') \\
		&= \frac{1}{2}[\Pr(D_{A,i}''=D_{A,i}|~\mathcal{B} = \mathcal{B}') + \Pr(D_{A,i}''=D_{A,i}|~\mathcal{B} \neq \mathcal{B}')] \\
		&=  \frac{1}{2}[\Pr(D_{A,i}''=D_{A,i}|~\mathcal{B} = \mathcal{B}',~D_{A,i}=D_{A,i}') \Pr(D_{A,i}=D_{A,i}') + \\
		&~~~~~~~~~~~~~~ \Pr(D_{A,i}''=D_{A,i}|~\mathcal{B} = \mathcal{B}',~D_{A,i} \neq D_{A,i}') \Pr(D_{A,i} \neq D_{A,i}') +1/2]\\
		& = \frac{1}{2}\left[1 \times \frac{1}{2} + 0 \times \frac{1}{2} + \frac{1}{2}\right]=\frac{1}{2}.
		\end{split}
		\end{equation*}
		Hence Alice and Bob can detect this eavesdropping and terminate the protocol with probability $1-{2}^{-m}$, where $m$ is the number of decoy states. Furthermore, since Eve has no idea about the value of the parameter $\theta$ and the exact position of the qubits corresponding to the secret message $M$, so without the classical information from Alice, Eve can not get any useful information by measuring the qubits of $Q_A^5$ in some random basis.
		
		\item \textbf{Information leakage attack:} It refers to the information about the secret message obtained by analyzing the classical channels by Eve. In other words, it is a measure of the information which Eve can get from the classical channel. Since in the present protocol, no measurement outcome corresponding to the secret bits is discussed by the classical channel, therefore Eve can not get any secret information from the communications in the classical channel.
		
		\item \textbf{Trojan horse attack:} In the present protocol, only Alice prepares all the qubits required for secure communication, and then she sends these qubits to Bob at once. Therefore this protocol is a one-way quantum communication protocol and hence Eve can not adopt the Trojan horse attack strategy to get any information about $M$.
	\end{enumerate}
	We have shown that our proposed protocol is secure against all the above-discussed attacks as in each case the legitimate parties can detect the presence of Eve with non-negligible probability.
	
	In the following section, we study the performance of this protocol in a realistic noisy quantum computer and illustrate results from IBM Quantum Computer.
	
\section{Implementation in a noisy quantum device}
The operations in our proposed protocol can be broadly represented as $U_B U_{Channel} U_A$ where $U_A$ and $U_B$ are the operations at the two ends (Alice and Bob respectively), and $U_{Channel}$ captures the action of the channel. Since Bob should receive the exact bit sent by Alice, if $\ket{q}$ is the qubit sent by Alice, we expect that in an ideal (noiseless and absence of eavesdropper) scenario
\begin{equation}
\label{eq:protocol}
    U_B U_{Channel} U_A\ket{q} = \ket{q}.
\end{equation}

Now in an ideal scenario our protocol requires $U_B = U_A^{-1}$. If $U_{channel} \propto I$, then this requirement suffices. Without loss of generality, we consider $U_{Channel}=nI$, where $n \in \mathbb{Z^{+}}$. The scalar $n$ also captures the finite time duration of the channel.

In reality, the channel is usually noisy and is no longer $\propto I$. If $p_{error}$ is the probability of error, then the noisy channel can be represented as

\begin{equation}
    U_{Channel}^{noisy} = (1-p_{error})nI + p_{error}\sum_{i=1}^{n} I_{e_i},
\end{equation}

where $I_{e_i}$ is some noisy version of the $i^{th}$ identity gate. Note that $I_{e_i}$ may not be equal to $I_{e_j}$ for $i \neq j$, and it is possible that for some $i$, $I_{e_i} = I$, i.e., some of the $n$ identity gates may be noise-free as well.

In such a scenario, the ideal operation of Bob should be $U_B = (U_{Channel}^{noisy})^{-1} U_A^{-1}$. However, since the action of the noise is unknown, it is not possible for Bob to apply this required operation in a realistic scenario. Furthermore, our protocol requires the preparation of $U_{\theta}$ gate for $\theta \in \Theta$. In near-term devices, which are noisy, this technique can be a victim of calibration error, i.e., the applied operation maybe $U_{(\theta + \delta \theta)}$ for some small $\delta \theta \in \mathbb{R}$. The protocol will be subject to measurement error as well.

Here, we execute this protocol on the IBM Quantum Computer (Armonk device). We assume different lengths of the quantum channel (i.e., various values of the scalar $n$). As discussed before, noise in this device deviates the realization of the quantum channel from $U_{Channel}$ to $U_{Channel}^{noisy}$. We execute this protocol for different values of $\theta$ as well and show that the protocol is robust against various sources of errors and the integrity of the protocol can be guaranteed with minimum overhead in a noisy scenario as long as the time duration of the ideal channel (i.e., the value of $n$) is below a certain threshold.


\subsection{Equivalence with Bit Flip Channel}
Prior to further discussion on errors, we want to mention explicitly a property of this QSDC protocol. Unlike general error correction scheme, in this protocol, it is not of urgency to preserve the exact state that is being sent from Alice to Bob. The ultimate goal is to ensure that Bob receives the exact bit that Alice has sent him with high probability. In other words, suppose Alice wants to send a qubit $\ket{q}$ to Bob corresponding to a classical bit $q$. However, in a realistic scenario, if the noisy operations of Alice, Bob and the channel are $U_A'$, $U_B'$ and $U_{channel}'$ respectively, then instead of the required $U_B U_{Channel} U_A\ket{q}$, we obtain $U_B' U_{Channel}' U_A'\ket{q}$. We do not care how the transmitted state $\ket{q}$ is being tampered with by the errors as long as $\bra{q}U_B' U_C' U_A'\ket{q} > 1 - \epsilon$ for some small $\epsilon > 0$.

Furthermore, let $\ket{q}$ be the original qubit transmitted by Alice, whereas Bob received $\ket{q'}$ which may not be the same as the original transmitted message. However, since $q \in \{0,1\}$, when Bob measures $\ket{q'}$ in the $\{\ket{0},\ket{1}\}$ basis, he either receives $q$ or $q \oplus 1$. Therefore, although the underlying channel may incorporate any error to the transmitted qubit, it is eventually equivalent to a single bit flip. Therefore, the overhead required for the error induced by the channel is the overhead to correct bit-flip errors.

\subsection{Simulation of the protocol in IBM quantum device}
In this subsection, we compute our protocol in the IBM Quantum Computer. However, for this computation, we have ignored the authentication portion. Rather we have only computed the communication portion, i.e., for each message qubit $\ket{q}$, we have computed the operation $U_B U_{Channel}U_A\ket{q}$, and shown the action of noise on it. The effect of noise can be mitigated using error correction. We aim to use the minimum overhead for error correction, which we discuss in the following subsection, followed by the computation results henceforth.

\subsubsection{Overhead for error correction}
To account for the imperfection of the channel, it is necessary to introduce error correction. However, for this protocol, we intend to introduce the minimum possible resource for error correction. Classically, a $3$-bit repetition code is sufficient to correct a single bit flip error. The repetition code is, in general, not extendable to the quantum domain, since (i) errors on qubits are not simple bit flips \cite{gottesman1997stabilizer}, and (ii) No Cloning Theorem prohibits cloning of any arbitrary quantum state \cite{wootters1982single}. However, we have already argued that the effective error on this protocol is indeed a simple bit flip. Furthermore, the qubits transmitted by Alice are either $\ket{0}$ or $\ket{1}$. Therefore, No Cloning Theorem does not restrict the use of repetition code in this scenario. The use of a distance $3$ repetition code ensures that to send $N$ qubits through a noisy channel, a total of $3N$ qubits are sufficient for error-free transmission as long as the error probability is below a particular threshold, which we now elaborate.

A distance-$3$ repetition code fails when at least two errors occur on the codeword. Therefore, if $p_{err}$ is the probability of error, then we should have
\begin{center}
    $\begin{pmatrix}
    3\\
    2
    \end{pmatrix}p_{err}^2 < p_{err}$,
\end{center}
which yields $p_{err} < \frac{1}{3}$.

In the following subsection, we show empirically that the action of noise is similarly for any angle $\theta$ selected for this protocol. However, the time duration of the channel restricts the distance of the code. We have represented a noisy quantum channel as $U_{Channel}^{noisy}$. We show that for the usual time duration of an identity gate in the IBMQ device, a distance 3 repetition code can protect this protocol from error as long as $n < 350$. For higher values of $n$, the noise in the device will lead to more than one error on expectation, and larger distance codes will be required for error-free transmission.

\subsubsection{Results of simulation in IBM Quantum Device}

\begin{figure}[htb]
    \centering
    \includegraphics[clip, trim=0 0 10cm 0, scale=0.4]{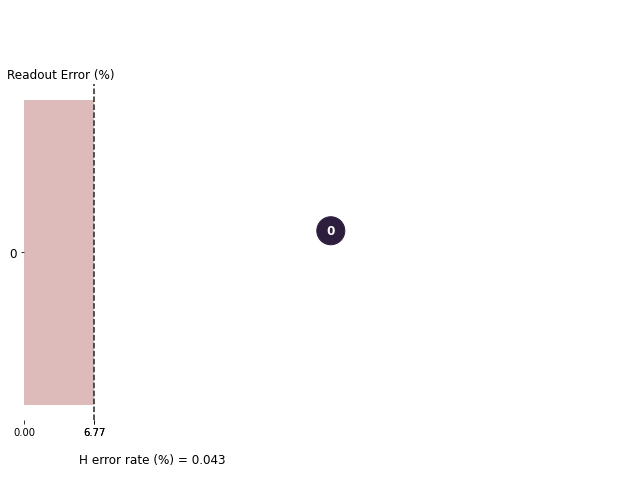}
    \caption{Specifications of the IBMQ Armonk quantum device as provided by IBM}
    \label{fig:armonk}
\end{figure}

In our protocol, once a $\theta$ is decided upon, each bit is encoded independently and sequentially by Alice. Similarly each qubit is decoded and measured independently and sequentially by Bob. Therefore, a single qubit quantum computer is sufficient to perform these operations. We have computed the encoding by Alice and the decoding by Bob, followed by measurement in the IBMQ Armonk device \cite{armonk} for various values of $\theta$ and various lengths ($n$) of the channel. IBMQ Armonk is a single qubit quantum computer with specifications shown in Fig.~\ref{fig:armonk}.

Computation on this device exposes our protocol to various device noise. Calibration error signifies the inaccuracy in the gate operation (denoted as H error rate in Fig.~\ref{fig:armonk}). Readout error, on the other hand, encapsulates the inaccuracy in measurement. If the measurement device is noisy, then it is possible that although the original output was $m$, due to measurement inaccuracy, it was noted down as $m \oplus 1$. Readout error is one of the most dominating sources of errors in current quantum devices (as shown in Fig.~\ref{fig:armonk} where the readout error rate is 6.7\% as compared to calibration error rate of 0.04\%). We shall discuss about the channel noise (particularly the $T_1$ error) later.

Qiskit \cite{Qiskit} has its own gate sets which are computed on their device. Such a gate is the $U3(\theta,\phi,\lambda)$ gate whose matrix form is
\begin{center}
    $U3(\theta,\phi,\lambda) = \begin{pmatrix}
    cos(\frac{\theta}{2}) & e^{-i \lambda}sin(\frac{\theta}{2})\\
    e^{i \phi}sin(\frac{\theta}{2}) & e^{i(\phi+\lambda)}cos(\frac{\theta}{2})
    \end{pmatrix}$,
\end{center}

where $0 \leq \theta, \phi, \lambda < 2\pi$ are the parameters. Different quantum gates can be generated by varying this parameter. Note that our required operation $U_{\theta} = U3(2\theta,0,0)$.

\subsubsection*{Effect of choice of angle}
\begin{figure}[htb]
    \centering
    \includegraphics[scale=0.5]{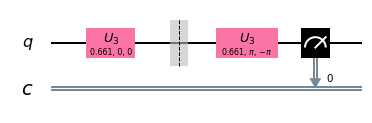}
    \caption{Circuit diagram of the QSDC protocol executed on the IBMQ Armonk device}
    \label{fig:circuit}
\end{figure}

First, we show the effect of the angle $\theta$ on the performance of the protocol in a realistic noisy scenario. For this portion, we do not consider the presence of channel. We have executed our protocol on the quantum device of Fig.~\ref{fig:armonk} for 20 equally spaced values of $\theta$ ranging from $0^\circ$ to $360^\circ$. We show the circuit for one such $\theta$ in Fig.~\ref{fig:circuit}. This figure shows the exact circuit that is being executed on the IBMQ Armonk device. The two gates are respectively the $U_{\theta}$ applied by Alice, and the $U_{\theta}^{-1}$ applied by Bob. Qiskit tends to optimize their circuit to reduce the execution overhead. Since we are applying two inverse operations sequentially, the optimization module of qiskit would lead to an identity operation. Therefore, we have forcefully introduced the barrier between the two gates which ensures that both the operations are executed as they are.

\begin{figure}[h!]
     \centering
     \begin{subfigure}[b]{0.47\textwidth}
         \centering
         \includegraphics[width=\textwidth]{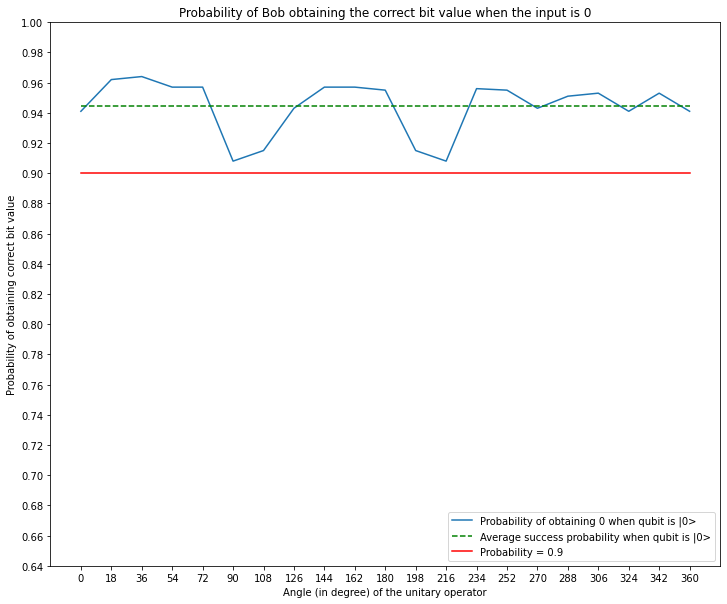}
         \caption{Performance when Alice sends $0$}
         \label{fig:send0}
     \end{subfigure}
     \hfill
     \begin{subfigure}[b]{0.47\textwidth}
         \centering
         \includegraphics[width=\textwidth]{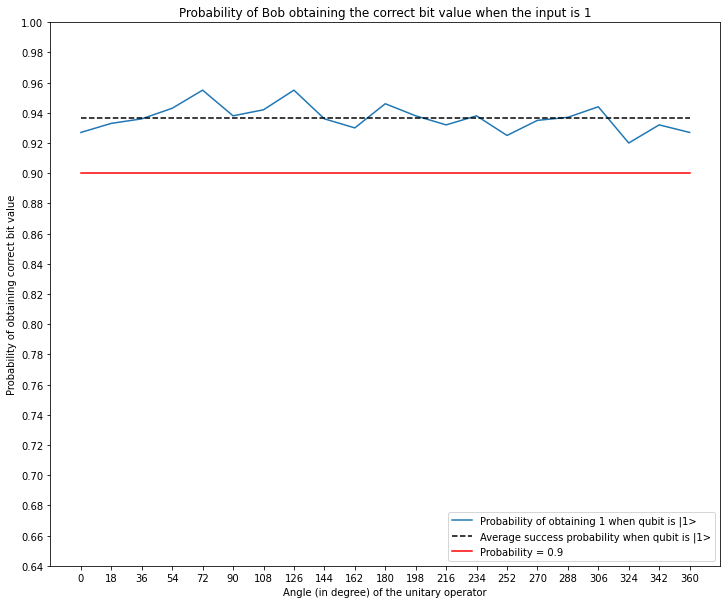}
         \caption{Performance when Alice sends $1$}
         \label{fig:send1}
     \end{subfigure}
     \caption{Action of noise in real quantum device}
\end{figure}

\begin{figure}[h!]
    \centering
    \includegraphics[scale=0.37]{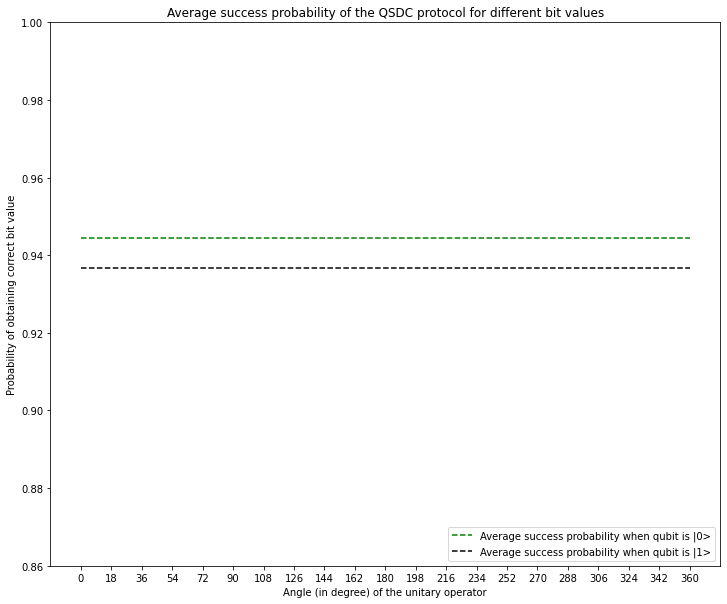}
    \caption{Average success probability for different bit values}
    \label{fig:avg}
\end{figure}

We have executed the protocol for the two scenarios - when the original bit is $0$ or $1$. Fig~\ref{fig:send0} and Fig.~\ref{fig:send1} shows the action of noise in real quantum device on the performance of the protocol. We see that Bob no longer obtains the original bit sent by Alice with certainty. However, it is evident from the figures that the choice of angle does not have any significant effect on the performance of the noisy protocol.

We note from Fig.~\ref{fig:avg} that the average performance is better when the qubit is $\ket{0}$ than when qubit is $\ket{1}$. This can be explained by the $T_1$ error. The natural tendency of any quantum state is to retain its lowest energy state ($\ket{0}$), or ground state. When a qubit is elevated to its excited state ($\ket{1}$), it has a natural tendency to release the excess energy to return to its ground state. This noise model \cite{nielsen2002quantum} is parameterized by $T_1$. In general, the probability that a qubit, prepared in the state $\ket{1}$, remains in that state after a certain time $t$ is given by
\begin{center}
    Prob($\ket{1}$) = $exp(-\frac{t}{T_1})$,
\end{center}

The qubits which are prepared in the state $\ket{1}$ are exposed to this error along with the other device noise. Therefore, naturally, the average probability of observing $\ket{1}$ is lower than that of $\ket{0}$. However, we note that for no value of $\theta$, the probability of correct transmission goes below 0.9.

\subsubsection*{Effect of the length of the channel}
\begin{figure}[htb]
     \centering
     \begin{subfigure}[b]{0.47\textwidth}
         \centering
         \includegraphics[width=\textwidth]{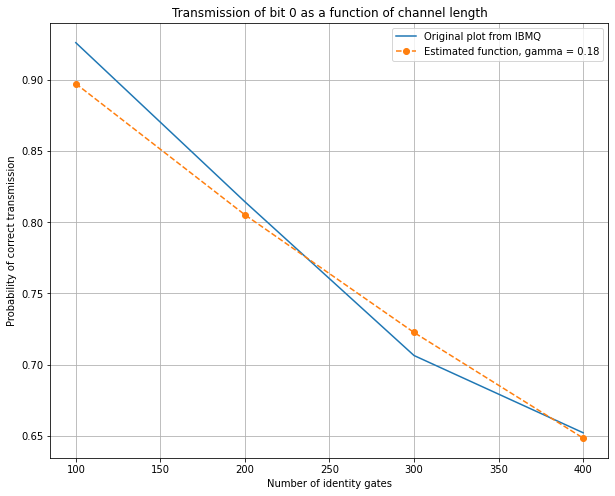}
         \caption{Performance variation with channel length when Alice sends $0$}
         \label{fig:channel0}
     \end{subfigure}
     \hfill
     \begin{subfigure}[b]{0.47\textwidth}
         \centering
         \includegraphics[width=\textwidth]{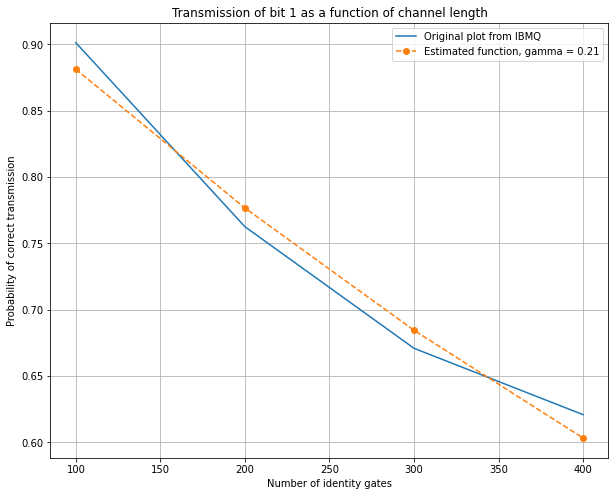}
         \caption{Performance variation with channel length when Alice sends $1$}
         \label{fig:channel1}
     \end{subfigure}
     \caption{Action of noise in real quantum device for different channel length}
\end{figure}

\begin{figure}[htb]
    \centering
    \includegraphics[scale=0.37]{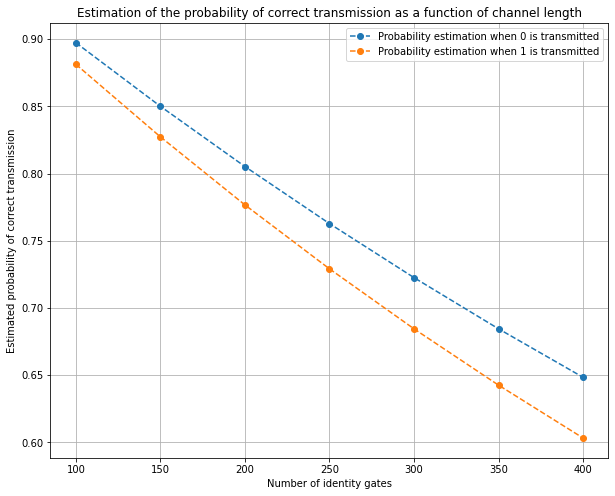}
    \caption{Estimated functions for success probability for varying channel length}
    \label{fig:estimation}
\end{figure}

Now, we incorporate the presence of a quantum channel. A quantum channel is not instantaneous. In order this simulate the finite time duration, we execute the circuit of Fig.~\ref{fig:circuit}, with $100 \leq n \leq 400$ identity gates in between the two $U_3$ operators. Each identity gate in the IBMQ Armonk device requires 142 ns to execute, and the error probability of each identity gate is $p_{error} = 0.001$. The probability that the channel remains error-free is $(1-p_{error})^n$. However, when we execute this circuit, it is subjected to other sources of errors apart from the channel noise only (e.g. calibration error, readout error). In order to account for these, we hypothesize that the probability of no error is
\begin{equation}
\label{eq:function}
    (1-p_{error})^{\gamma n},
\end{equation}

for some scalar $\gamma$. In Fig.~\ref{fig:channel0} and ~\ref{fig:channel1}, we show the probability of correct transmission as a function of the length of the channel. We estimate the value of $\gamma$ in each case through curve fitting and observe $\gamma = 0.18$ for the transmission of bit 0, and $\gamma = 0.21$ for the transmission of bit 1. The estimated functions are plotted in Fig.~\ref{fig:estimation} to show a comparison of the variation in probability for the bits 0 and 1. We see that, similar to Fig.~\ref{fig:avg}, the transmission of 1 is more prone to error than that of 0. This can be similarly explained as before via the $T_1$ error. This is, in fact, the reason for obtaining two different values of $\gamma$ for the two bits.

We have already argued that a distance 3 repetition code is applicable for correcting errors only when the probability of no error is $\geq \frac{2}{3} = 0.66$. We note from Fig.~\ref{fig:estimation} that when the number of identity gates is $\sim 350$, the estimated success probability of both 0 and 1 goes below the required threshold. Therefore, in order to use the minimum overhead of 3 qubit repetitions, it is necessary that the channel length is $< 350$ identity gates. Nevertheless, in case the channel length is greater, then higher distance repetition codes can be used for error-free transmission.

\section{Conclusion}\label{conclusion}
	In this paper, we propose a QSDC protocol with user authentication using single qubits prepared on a randomly chosen arbitrary basis. In this protocol, before starting the communication process, Alice and Bob share their secret identities through a secure QKD to authenticate each other. In the proposed QSDC protocol, Alice, the message sender, prepares all the single qubits and sends them to the receiver Bob, i.e., this is a one-step one-way quantum communication protocol. After receiving the qubits, Bob only performs measurement and applies unitary operations to the received particles to get the secret message of Alice. Moreover, the present protocol does not use entanglement as a resource. We discuss the security of the protocol and show that our proposed protocol defeats all the familiar attack strategy and the eavesdropper could not get on any information about the secret message. The curse of executing such protocols in near-term devices is that they become susceptible to noise in the device. We have computed the protocol in the IBMQ Armonk device which is a single qubit device, and therefore perfectly captures the sequential structure of the protocol. We find that our protocol is quite robust to error, and a simple distance $3$ repetition code is sufficient for reliable transmission as long as the length of the quantum channel is less than $350$ identity gates. Therefore, in order to transmit $N$ qubits in such a noisy scenario, $3N$ qubits are sufficient, and it does not require any complex gate operations for preparing logical qubits as well.
	\bibliographystyle{unsrt}
	\bibliography{main}
\end{document}